# Towards Electrophysiological and Histological Mapping of Upper Limb Nerves in Pigs Using Epineural Stimulation

Baum J[1], Chamot-Nonin M[2], Oppelt V[3], Guiraud D[2], Azevedo Coste C[1], Guiho T[1]

[1]INRIA, University of Montpellier, France
[2]NEURINNOV company, Montpellier, France
[3]CorTec company, Freiburg, Germany

**Abstract:** *Understanding the relationship between nerve anatomy and the functional outcomes of electrical stimulation is critical for optimizing neural interface design. In this study, we conducted acute experiments on four pigs in which epineural cuff electrodes with multiple contacts were placed around upper limb nerves. A subset of electrical stimulation configurations - previously identified via computational study - was applied, and the resulting evoked electromyographic (EMG) responses were recorded from target muscles. Muscle recruitment curves were extracted and analysed offline to quantify activation patterns. Following the electrophysiological experiments, the stimulated nerves were harvested and processed for histological analysis to visualize fascicular organization and distribution. This work presents preliminary results from the combined analysis of muscle activation profiles and fascicle anatomy in one animal. Our findings aim to inform the design of stimulation strategies by linking electrode configuration to selective muscle recruitment, ultimately contributing to more effective neuromodulation and neuroprosthetic applications.*

## Introduction

Implanted peripheral nerve stimulation has shown promise in restoring impaired motor functions of the upper limb—particularly hand and wrist movements—in individuals with complete spinal cord injury [1]. Compared to epimysial or intraneural electrodes, the use of multicontact epineural electrodes offers a less invasive alternative while maintaining access to multiple motor pathways [2]. However, achieving selective activation of specific muscle subgroups to produce functional movements remains a significant challenge. Currently, stimulation parameters and electrode configuration are typically adjusted empirically, based on observed motor responses, which limits reproducibility and precision. This approach is further complicated by inter- and intraindividual variability in nerve morphology, particularly regarding the complex fascicular organization and plexiform architecture of the human brachial plexus [3]. Understanding how these anatomical factors influence stimulation outcomes is essential for optimizing electrode design and stimulation strategies.

To partially address this scientific challenge, equivalent electrical models of axons were developed to enable the in-silico identification of the most selective and efficient current configurations. Cuff epineural electrodes were considered in the model due to their moderately invasive characteristics and widespread commercial availability - in contrast to other architectures such as TIME intraneural electrodes which might be less suitable for clinical trials due to mechanical constraints associated with their transversal positioning through the nerves. Previously, subset of stimulation configurations identified as selective by the model [4] was evaluated in 28-days clinical trials investigating peripheral nerve stimulation for hand and wrist rehabilitation [5]. These configurations were further refined empirically using evoked electromyography (eEMG) and observation of evoked movements to elicit specific functional gestures. However, this empirical approach is time-consuming, and only a limited portion of the extensive parameter space can be explored clinically for the refinement of epineural functional electrical stimulation (FES). Optimization of stimulation parameters and electrode configurations should be more comprehensive and objective, while remaining time-realistic. The development of automated tools capable of correlating epineural current patterns with evoked electromyographic (eEMG) responses is essential to systematically explore the extensive parameter space and enhance individualized epineural stimulation. However, existing models are either based on stereotypical assumptions of nerve structure [6], or on histological observations that lack electrophysiological validation [7]. A recent study attempted to bridge this gap by combining anatomical and electrophysiological data using a specific epineural electrode [8], but the stimulation paradigm was limited to bipolar configurations. While this work highlights the value of building subject-specific models, it remains constrained in selectivity. This challenge is then compounded by inter-individual variability, which further complicates the development of accurate and generalizable models. With these considerations in mind, we designed a preclinical study that aimed at elucidating the relationship between upper limb nerve architecture and electrophysiologically evoked responses. To enhance the translational relevance of the findings, priority was given to a large animal species. In particular, the selected species – the domestic pig (*sus scrofa domesticus*) - exhibits anatomical similarities to humans, notably in the distribution of the brachial plexus, including the median, radial, and ulnar nerves, which are functionally conserved in controlling wrist and finger flexion and extension. [9].

In this article, on top of confirming the relevance of swine for electrophysiological studies of upper limbs functions, we exploited eEMGs to draw a parallel between evoked muscular activities elicited by a subset of presumably selective multipolar stimulation configuration/parameters and actual nerve anatomy - obtained via histological analyses of stimulated nerves.

## Material and Method

After approval by the animal ethics committee of Languedoc-Roussillon (France, Agreement number #47593-2024021614231926v2), acute pre-clinical trials were conducted on four neurotypical juvenal swine (2-3 months age, 30-40 kg).

### Surgery protocol

After premedication, anaesthesia was induced with propofol and maintained via intravenous infusion of propofol mixed with sufentanil for analgesia. Swine condition was then monitored throughout the procedure (body temperature; heart rate; blood saturation; breathing rate with air and oxygen mechanically assisted ventilation).

After ensuring proper anaesthesia, swine axillary fossa was incised to expose the brachial plexus. After anatomical prospection, the median nerve was further identified with a bipolar stimulation stick inducing contraction of predefined forelimb muscles, namely "Flexor carpi radialis" (FCR), "Flexor digitorum superficialis" (FDS), "Pronator teres" (PT). "Extensor carpi radialis" (ECR) muscle was also identified to enable monitoring of antagonist contractions. A multichannel cuff electrode (CorTec™, Germany) with two plain contacts on distal and proximal extremities and six individual contacts distributed along the central circumference (Fig. 1, see [2] for further details) was then wrapped around the nerve before connection to a 16 channels bench-top neural stimulator (SimENS, Neurinnov's®, France) for electrophysiological measurements. Armpit incision was then filled with saline to enable moisturization of the nerve and electrical continuity between the electrode contacts and the nerve.

### Stimulation and Electrophysiological recordings protocol

Epineural stimulation was delivered using asymmetric biphasic pulses - 150 μs phase width - at a frequency of 35 Hz. Model-based preidentified stimulation configurations were investigated, ie. rings and "steering current" (STR). The STR configuration is characterized by the distribution of current between one cathode and three anodes leading to anodic current repartition of 1/3 on each ring and on the inner contact opposite to the cathodic one. STR are numbered from 1 to 6 according to cathodic central contact position around nerve circumference (Fig. 1, [2]).

For each configuration (7 in total), stimulation intensity was initially set at a low value (few tens of μV) before gradually increasing - step of 9 μA - until estimation of saturation of the muscle recruitment. Each step duration was 4.5 s long leading to trains of 19 stimulation pulses and 19 M-waves per intensity.

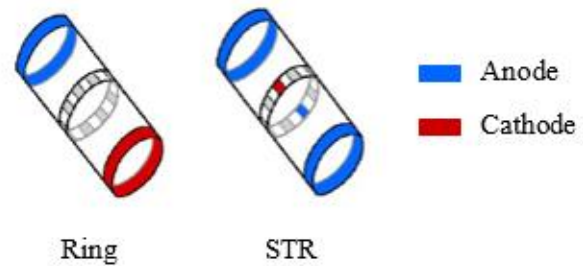

Figure 1: Schematic representation of ring and STR configurations. Adapted from [2]

Targeted muscles were localized via skin palpation further confirmed through stimulation-evoked contractions and movements. Pairs of Teflon-insulated stainless-steel wires were inserted percutaneously to record evoked EMGs (eEMGs). The signals were then amplified with a gain of 5000, band pass filtered between 1 Hz and 5 kHz (Butterworth filter of 4th order) et notch filtered at 50 Hz using a g.Bsamp GT201 amplifier (g.tec medical engineering, Austria) before sampling at 20 Khz (PowerLab 16/35, AD Instrument, New Zealand). The associated LabChart software (LabChart 7, AD Instrument, New Zealand) was used throughout the experiments to record and also stream signals in real time

### Histological protocol

Following euthanasia of the swines, both the stimulated and control nerves were harvested and immediately immersed in saline solution. The positions of the electrode contacts were marked on the epineurium using tattoo ink prior to nerve sectioning. Nerve fragments were fixed in 10% paraformaldehyde (PFA) for 48 hours, followed by dehydration through a graded ethanol series prior to paraffin embedding. The paraffin-embedded nerve fragments were sectioned at a thickness of 3 μm and subsequently stained with hematoxylin-eosin-saffron (HES) to visualize intraneural structures, or processed for choline acetyltransferase (ChAT) immunohistochemistry to specifically evaluate the distribution of motor fibers within the fascicles.

### eEMGs processing

Raw recorded eEMGs were further band pass filtered between 10 and 500 Hz (Butterworth filter of 4th order). For each muscle and each stimulation intensity, repeated M-waves were averaged before peak-to-peak calculation. The obtained data were then normalized using maximum eEMG peak-to-peak values (obtained over all recordings) before plotting recruitment curves.

## Results

In this section only preliminary results from pig #2 (P#2) are presented.

**Relevance of swine's nerve model for an electrophysiological study of upper limbs innervation.**

As shown in Fig. 2, swine's eEMGs provided recruitment curves indicating selective stimulation of forelimb muscles Recruitments show selectivity for FCR; PT; and ECR for STR2 configuration- ie. with the second central active site as cathode - at 225µA, whereas opposite contact (STR5) shows selectivity for FDS muscle at 200µA.

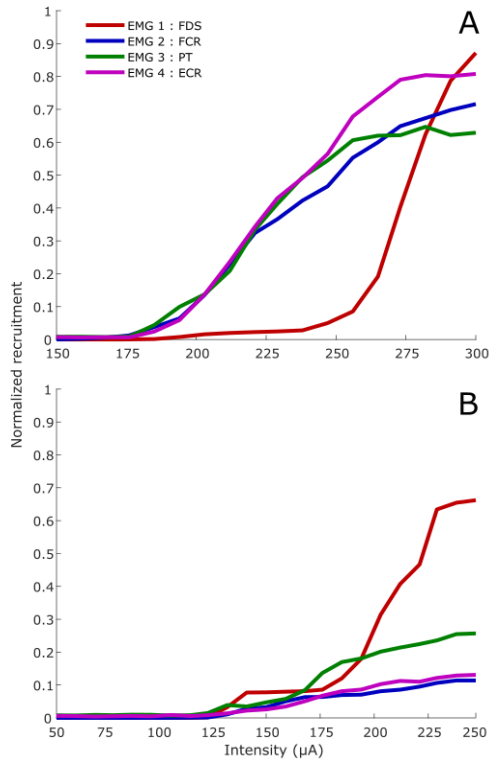

Figure 2: Normalized recruitment curves from P#2's eEMGs in STR2 (A) and STR5 (B) stimulation configurations.

Histological sections of the median nerve reveal its fascicular organization and show slight variations of fascicle number and position over short distances - ie 100 µm (distance between two successive histological cross-sections). On top of that, fascicle's size also varies between sections. In Fig. 3, section A shows 18 fascicles against 19 in section B. This last has two fascicles resulting from division of the fascicle observed in the previous section (phenomenon highlighted by frames C).

**Recruitment compared with nerve's histology**

Polar representation relative to each STR configuration provides information about recruitment evolution against increases in stimulation intensity. As shown in Fig. 4, FCR, PT, and ECR muscles were recruited preferably by STR 2 and 3, with maximum values obtained for STR 3. Concerning FDS, recruitment was only achievable with STR 3, 4, and 5. STR 4 seems to be the best configuration for early recruitments, whereas STR 3 generates the highest recruitment values for FDS.

Regarding histological analysis, anti-ChAT immunohistochemistry revealed motor fiber distribution among fascicles; suggesting that some fascicles are more likely to participate to motor control than others due to their higher concentration of motor fibers (Fig. 5)

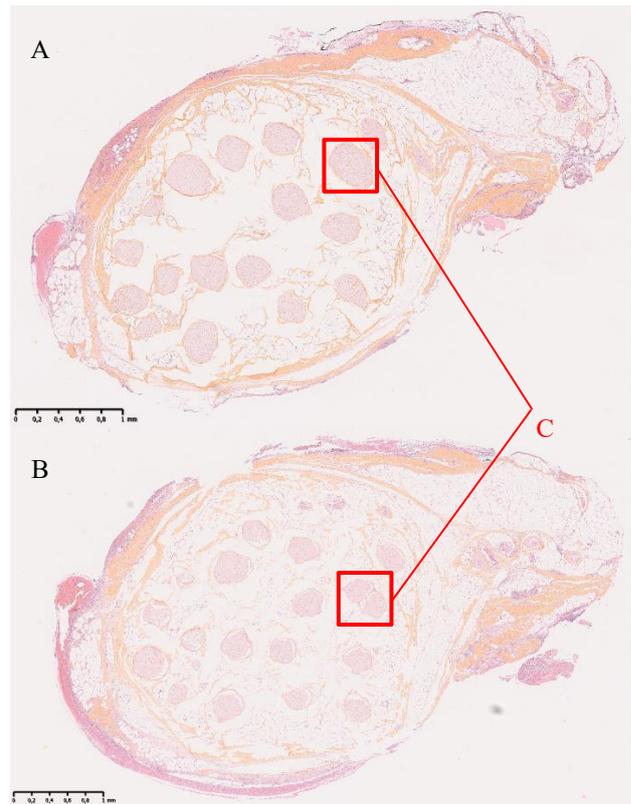

Figure 3: P#2 median nerve transversal sections stained with HES. A (the most proximal) and B sections are taken from the same nerve and are 400µm apart. C frames indicate the same fascicle (dividing in section B).

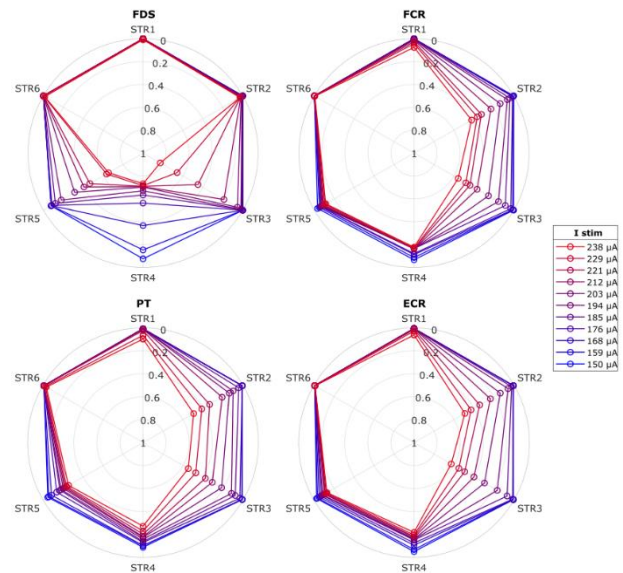

Figure 4: Polar representation of muscle recruitments from P#2 depending on STR configuration selected. Stimulation intensities are increasing from 150 to 238µA (blue to red colour lines) for each muscle. Recruitment values are normalized, from 0 on circumference of polar plot to 1 on the centre.

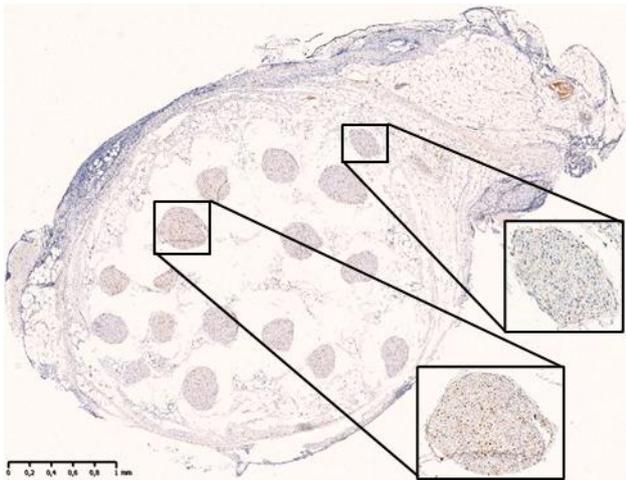

Figure 5: Anti-ChAT immunochemistry staining on P#2 median nerve transversal section. Motor fibres appear coloured in brown. Blue stains are cell nucleus of conjunctive tissues in nerve.

## Discussion

Recruitment curves are an efficient way to assess stimulation selectivity. Their analysis allows for refinement of stimulation configuration supposedly resulting in a wider range of functional movements. Plexiform organisation of brachial nerves observed in pigs – and somewhat similar to humans - has still to be confirmed on more individuals and longer transversal distance. This gap will be filled in the next few weeks after receipt and analysis of new histological sections from additional nerve samples.

On the other hand, anti-ChAT immunochemistry revealed the location of the fascicles containing most of the motor fibers. Combining those physiological parameters to polar representation could provide a better understanding of optimized stimulation configurations although this will require further analysis - especially regarding correlation of recruitment and location of motor axons within the nerve.

This work contributes to a better understanding of how stimulation parameters influence recruitment. It also raises the question of whether EMG-based recruitment data alone can reliably infer internal nerve architecture. Ultimately, the integration of anatomical and functional data may enhance computational models and guide the design of more selective and efficient neuroprosthetic systems.

## Conclusions

Regarding the obtained recruitment curves, electrical stimulation configurations determined through computational models appear to elicit selective muscle activation in swine upper-limb. Moreover, histological analysis reveals the actual distribution of motor fibers within the nerve, information that should ultimately contribute to the validation/improvement of the current computational models of peripheral nerve stimulation. We are planning to consolidate these preliminary results by pursuing the analyses of recruitment curves and histological data from the remaining animals. Results will be presented at the conference.

## Acknowledgement

The authors thank the RAM-PTNIM (Nîmes, France,) facilities for supporting with animal caring and surgical logistic and the RHEM (Montpellier, France) for advices on nerve sampling and, histological analysis and scanning of microscope slides.

Funded by AI-Hand European project (EIC Pathfinder – Grant No. 101099916) and Plasticistim (EMERGENCE program - No #23004566, Région Occitanie, France).

## References

[1] K. L. Kilgore et al., « Evolution of Neuroprosthetic Approaches to Restoration of Upper Extremity Function in Spinal Cord Injury », Top Spinal Cord Inj Rehabil, vol. 24, no 3, p. 252-264, 2018, doi: 10.1310/sci2403-252.

[2] C. A. Coste et al., « Activating effective functional hand movements in individuals with complete tetraplegia through neural stimulation », Sci Rep, vol. 12, no 1, p. 16189, oct. 2022, doi: 10.1038/s41598-022-19906-x.

[3] I. Delgado-Martínez, J. Badia, A. Pascual-Font, A. Rodríguez-Baeza, et X. Navarro, « Fascicular Topography of the Human Median Nerve for Neuroprosthetic Surgery », Front. Neurosci., vol. 10, juill. 2016, doi: 10.3389/fnins.2016.00286.

[4] M. Dali et al., « Relevance of selective neural stimulation with a multicontact cuff electrode using multicriteria analysis », PLOS ONE, vol. 14, no 7, p. e0219079, juill. 2019, doi: 10.1371/journal.pone.0219079.

[5] C. Fattal et al., « Restoring Hand Functions in People with Tetraplegia through Multi-Contact, Fascicular, and Auto-Pilot Stimulation: A Proof-of-Concept Demonstration », J Neurotrauma, vol. 39, no 9-10, p. 627-638, mai 2022, doi: 10.1089/neu.2021.0381.

[6] K. E. I. Deurloo, J. Holsheimer, et P. Bergveld, « Fascicular Selectivity in Transverse Stimulation with a Nerve Cuff Electrode: A Theoretical Approach », Neuromodulation: Technology at the Neural Interface, doi: 10.1046/j.1525-1403.2003.03034.x.

[7] D. K. Leventhal et D. M. Durand, « Subfascicle stimulation selectivity with the flat interface nerve electrode », Ann Biomed Eng, vol. 31, no 6, p. 643-652, juin 2003, doi: 10.1114/1.1569266.

[8] S. L. Blanz et al., « Spatially selective stimulation of the pig vagus nerve to modulate target effect versus side effect », J. Neural Eng., vol. 20, no 1, p. 016051, févr. 2023, doi: 10.1088/1741-2552/acb3fd.

[9] A. S. Hanna et al., « Brachial plexus anatomy in the miniature swine as compared to human », J Anat, vol. 240, no 1, p. 172-181, janv. 2022, doi: 10.1111/joa.13525.

## Author's Address

BAUM Jonathan - jonathan.baum@inria.fr